\documentclass[twocolumn,pre,showpacs,amsmath,amssymb,floatfix]{revtex4}
\usepackage{graphicx}
\usepackage{dcolumn}
\usepackage{bm}

\begin{document}

\title{Structural phase transition in evolving networks}

\author{Sang-Woo Kim and Jae Dong Noh}
\affiliation{Department of Physics, University of Seoul,
  Seoul 130-743, Korea}

\date{\today}
\begin{abstract}
A network as a substrate for dynamic processes
may have its own dynamics. 
We propose a model for networks which evolve together 
with diffusing particles through a coupled dynamics, and investigate emerging 
structural property. 
The model consists of an undirected weighted network of fixed
mean degree and randomly diffusing particles of fixed density.
The weight $w$ of an edge increases by the amount of traffics
through its connecting nodes or decreases by a constant factor.
Edges are removed with the probability $P_{rew.}=1/(1+w)$ 
and replaced by new ones having $w=0$ at random locations. We find that
the model exhibits
a structural phase transition between the homogeneous phase characterized by
an exponentially decaying degree distribution and the
heterogeneous phase characterized by the presence of hubs. 
The hubs emerge as a consequence
of a positive feedback between the particle and the edge dynamics.
\end{abstract}
\pacs{89.75.Hc, 05.70.Fh, 05.40.-a, 64.60.Ht}
\maketitle

Complex networks have been the subject of extensive researches 
for the last decade.
They have a heterogeneous structure, which makes them distinct from 
the periodic lattice and the random network. 
Researches have been focused on characterizing the structure and revealing
the mechanism leading to it~\cite{Watts98,Albert02,Newman03,Dorogovtsev02}.
Some complex networks play the role of a substrate on which 
dynamic processes, either equilibrium or nonequilibrium, take place. 
Implication of the structural heterogeneity on the dynamic processes has 
also attracted a lot of interests~\cite{Boccaletti06,Dorogovtsev08}.

In most studies dynamics of a network itself and additional 
degrees of freedom on it are treated separately.
These approaches are meaningful when characteristic time scales associated 
with each of them are completely separated. When they are comparable, 
one needs to consider 
the dynamics of both kinds of degrees of freedom simultaneously.
Along these lines, dynamic models for a social network coupled with 
a game-theoretical dynamics or an opinion dynamics were studied in 
Refs.~\cite{Ebel02,Zimmermann04,Gil06,Holme06}. 

The coupled dynamics was also studied in a dynamic model for
a transportation or an information network by the present authors in 
Ref.~\cite{SWKim08}. This study was motivated by the synaptic plasticity 
in neural networks~\cite{Lomo03}. Synaptic links in a neural network
may strengthen or weaken depending on synaptic activities, which results in 
a plastic deformation of a network. In the model~\cite{SWKim08}, 
particles diffuse over a network and edges are rewired at the 
rate depending on particle flows in such a way that edges contributing more 
to transport are more robust.
It was found that the coupled dynamics leads to an instability toward the 
formation of a hub. Although the model is useful in studying the dynamical 
origin for the emergence of a hub, it lacks a parameter with which one can 
control the strength of the instability.

In this paper we consider a model as an extension of the study 
in Ref.~\cite{SWKim08}.
The model consists of an undirected weighted network 
and diffusing particles. 
There are $N$ nodes with the mean degree $\langle k \rangle$ and 
particles with the density $\rho$. An edge $e$ is assigned to a weight
$w_e\geq 0$, and a node can accommodate multiple particles. The edges,
weights, and particles evolve in time
as follows: At each time step, every particle hops independently to a 
neighboring node selected at random. Whenever a node is reached by a
particle, the weight of all edges attached to it is increased by unity.
Then, with the probability $p_{rew.}=1/(1+w_e)$, each edge $e$ is removed
and replaced by a new one with $w=0$  
between a pair of nodes selected randomly. 
As a regularization procedure, the weight of all edges are degraded by the 
factor $r$, i.e., $w_e \rightarrow (1-r)w_e$ for all $e$.

This model allows one to study the emerging property of a complex network  
evolving through a coupled dynamics with a transport system. For
simplicity, we adopt the system of noninteracting random walkers as a
transport system. Each edge is assigned to the weight which measures the amount 
of traffics handled by connecting nodes.
The random walkers move along edges, while edges are rewired at the rate which
is a decreasing function of the weight. That is to say, the more 
contribution to the traffic an edge makes, the more robust it is.

When the degradation factor $r$ is zero, the model reduces to the one studied
in Ref.~\cite{SWKim08}. It was shown that the coupled dynamics between 
the edges and particles leads to an instability toward the formation of 
a hub. Interestingly, the hub emerges in the two distinct ways.
When the particle density is low, the hub appears spontaneously 
after overcoming a dynamic barrier via statistical fluctuations. 
On the other hand, when the particle density is high, nodes compete 
for edges and one of them survive as a hub eventually. 
In the original model the weight of an edge can increase indefinitely, which
may not be the case when an edge loses its strength due to e.g. ageing. 
The current model incorporates such an effect by introducing the degradation 
factor $r$. We will show that the model with the degradation factor
exhibits richer behaviors. 

The degradation factor $r$ limits the growth of the edge weight. Hence the
model can have a stationary state with nonzero values of $r$. 
Furthermore, there may be a phase transition. When $r$ is large, one
expects that edges are rewired at so high rates that the network remains 
like a random network without any hub. In the opposite
case with small $r$, edges attached to a certain node may 
become robust by gaining larger weights. Such a node can grow into a hub
as in the case with $r=0$. Practically, we define the hub as a node whose
degree scales algebraically with the total number of nodes.

We have performed numerical Monte Carlo simulations 
in order to examine whether there is
a transition between the states with and without hubs. 
We present detailed results of numerical simulation studies.
Initially we start with a random network over which particles are
distributed randomly. The mean degree is fixed to $\langle k\rangle=4$.
Then we study the time evolution and the stationary state property 
of the system.

\begin{figure}[t]
\includegraphics*[width=\columnwidth]{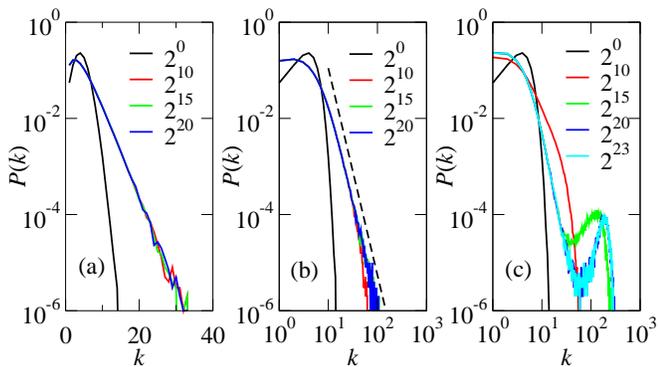}
\caption{Time evolution of the degree distribution at 
$r=0.20$ (a), $0.055$ (b), and $0.01$ (c). 
The network size is $N=10^3$ and the particle density is
$\rho=1.0$. Each data set is obtained by averaging over $N_S=10^3$
samples. The dashed line in (b) has a slope $-4.3$.}\label{fig1}
\end{figure}

The degree distribution has been measured as one varies $r$ with fixed 
$\rho=1$, which is presented in Fig.~\ref{fig1}. The degree distribution
function $P(k)$ is defined as the fraction of nodes having $k$ edges.
At $r=0.2$, the degree distribution in the stationary state aligns along a
straight line in the semi-log plot~(see Fig.~\ref{fig1}(a)). 
This means that the degree distribution
follows an exponential decay as 
$P(k)\sim e^{-k/k_0}$.
We observe a distinct feature at $r=0.01$. 
While the degree distribution decays at small 
values of $k$, there appears a peak in the large $k$ region. 
We will show later that both the number of nodes contributing to the peak
and the degree of them scale algebraically with the network size $N$,
respectively. Namely, the peak signals the emergence of multiple hubs.
In the intermediate case with $r=0.055$, the degree
distribution follows a power law 
\begin{equation}
P(k) \sim k^{-\gamma}
\end{equation} 
with the exponent $\gamma\simeq 4.3$.

Numerical data presented in Fig.~\ref{fig1} show that the model undergoes a
structural phase transition between the stationary states 
with and without hubs. Such phases will be denoted as a {\em heterogeneous} 
phase and a {\em homogeneous} phase, respectively.
The transition point can be estimated accurately from the effective exponent
defined as
\begin{equation}
\gamma_{eff}(k) = -  \ln[ P(ak) / P(k)] / \ln a
\end{equation}
with a constant $a=2$. As a function of $k$, it will grow without bound 
if the degree distribution decays exponentially. In the presence of
the peak for hubs, the effective exponent will be a non-monotonic function
of $k$.
If the degree distribution follows asymptotically a power law as 
$P(k)\sim k^{-\gamma}$, the effective exponent will converge to $\gamma$.

\begin{figure}[t]
\includegraphics*[width=\columnwidth]{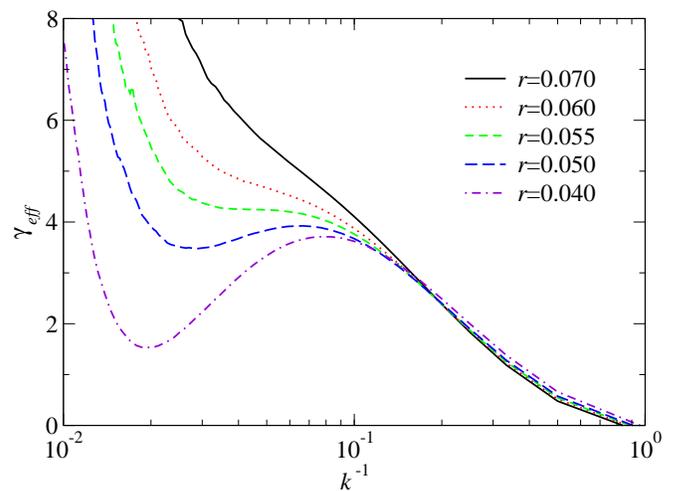}
\caption{Plots of the effective exponent $\gamma_{eff}(k)$ for the degree
distribution of networks with $N=1000$ and $\rho=1.0$.
}\label{fig2}
\end{figure}

\begin{figure}[t]
\includegraphics*[width=\columnwidth]{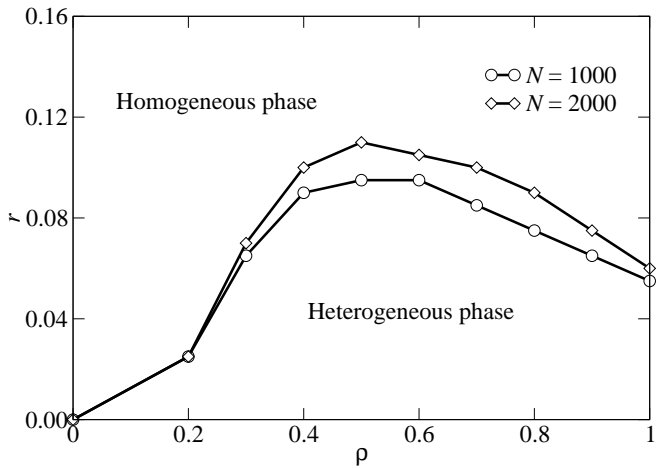}
\caption{Numerical phase diagram obtained from simulations with $N=1000$ and
$2000$.}\label{fig3}
\end{figure}

In Fig.~\ref{fig2}, we present the plot of the effective exponent at several
values of $r$ at $\rho=1.0$. When $r=0.055$, there appears a plateau at
$\gamma_{eff} \simeq 4.3$. Above and below $r=0.055$, the effective exponent
plot shows the characteristic of the exponential decay and the peak for
hubs, respectively. Hence we conclude that
the structural phase transition takes place at $r=r_c = 0.055(5)$. 
The degree distribution at $r=r_c$ follows the power-law decay 
with the exponent $\gamma \simeq 4.3$. 
Note that the blowup of $\gamma_{eff}$
for $k^{-1}\lesssim 0.02$ is due to a finite size effect.
Repeating this analysis at other values of $r$ and $N$, we obtain 
the numerical phase diagram as shown in Fig.~\ref{fig3}. Although a 
finite size effect is rather large up to $N=2000$, the numerical phase
diagram show a clear evidence for the phase transition.
The degree exponent remains almost constant along the phase boundary.

We present an analytic theory which explains the mechanism of the phase
transition. In order to describe the dynamics, 
one needs to consider the degrees of
all nodes, the weights of all edges, and the particle occupation number at
all nodes. It is difficult to consider the whole dynamics, so we develop 
an approximate theory as below.

Consider an arbitrary node $i_0$. The degree of the node 
at time $t$ will be denoted by $K(t)$. We assume that there exists a 
characteristic value of the weight $\Omega(t)$ for the $K$ edges of $i_0$. 
We make a further assumption that the other part 
can be regarded as a uniform medium where edges are rewired at a constant 
rate $s$. For these assumptions, our description is a mean field theory,
which works only when structural heterogeneity of the 
system is negligible. It is not valid in the heterogeneous phase with
hubs since it loses self-consistency. 
Nevertheless, we can learn when and why the structural phase transition 
will occur from the breakdown of self-consistency.

An edge gains a weight when particles arrive at its connecting nodes.
Diffusing particles on complex networks reach the stationary 
state very rapidly~\cite{Noh04,Noh07}. So we adopt a quasi-stationary state 
assumption that the particle distribution is approximated by the 
the stationary state distribution to a given network at each moment.
The stationary state particle 
distribution function is strictly proportional to the degree~\cite{Noh04}. 
The quasi-stationary state assumption, which was also made in
Ref.~\cite{SWKim08}, allows us to integrate out the particle degrees of
freedom. Then, the rate equation for the
weight variable $\Omega$ in time-continuum limit is given by
\begin{equation}\label{dot_Omega}
\frac{d\Omega}{dt} = \frac{\rho K}{\langle k \rangle} + \rho - r \Omega\ .
\end{equation}
The first two terms account for the gain coming from the visit of particles
to the node $i_0$ and its neighboring node, respectively.
The last term accounts for the loss due to the degradation. 

The degree variable $K(t)$ follows the rate equation
\begin{equation}\label{dot_K}
\frac{dK}{dt} = s - \frac{K}{1+\Omega} \ .
\end{equation}
The first term accounts for the attachment of a randomly rewired edge to
$i_0$, and the second term accounts for the rewiring of each of the $K$ 
edges with the probability $1/(1+\Omega)$. 

The flow governed by Eqs.~(\ref{dot_Omega}) and (\ref{dot_K}) has an
attracting fixed point when $r > r_c$ with
\begin{equation}\label{r_c}
r_c = \frac{s \rho }{\langle k \rangle} \ .
\end{equation}
The fixed point is located at 
\begin{equation}
K_0 = \frac{s(r+\rho)}{(r-r_c)} \ , \ 
\Omega_0 = \frac{(\rho+r_c)}{(r-r_c)}  \ .
\end{equation}
Irrespective of an initial condition, the flow converges to the fixed point.

\begin{figure}[t]
\includegraphics*[width=\columnwidth]{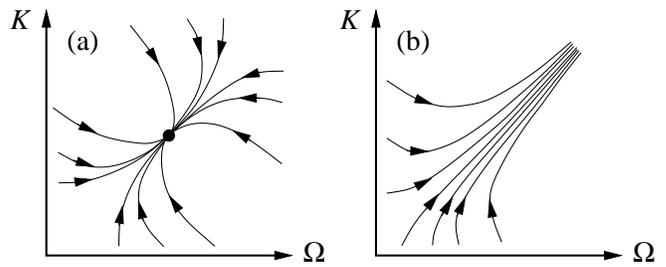}
\caption{Schematic flow diagram in the $(\Omega,K)$ plane when
$r<r_c$ (a) and $r>r_c$ (b). The fixed point $(\Omega_0,K_0)$ is represented
with a filled circle in (a).}\label{fig4}
\end{figure}

When $r<r_c$, there does not exist an attracting fixed point at finite
values of $K$ and $\Omega$. They grow unboundedly. 
The blowup solution invalidates the assumption that the network remains
homogeneous. It signals the emergence of a hub. 
The flow pattern is sketched schematically in Fig.~\ref{fig4}.

The mean field theory confirms that the structural phase transition indeed 
takes place. It also reveals the mechanism for 
the emergence of the hub. Following Eq.~(\ref{dot_Omega}), an increase of
$K$ accelerates the growth of $\Omega$. Likewise, an increase of $\Omega$
accelerates the growth of $K$. This shows that there is a positive feedback
between $K$ and $\Omega$. 
Actually the edge weight growth is driven by diffusing particles. 
Therefore, we conclude that the coupled dynamics of the network and
diffusing particles can lead to the heterogeneous network structure.

\begin{figure}[t]
\includegraphics*[width=\columnwidth]{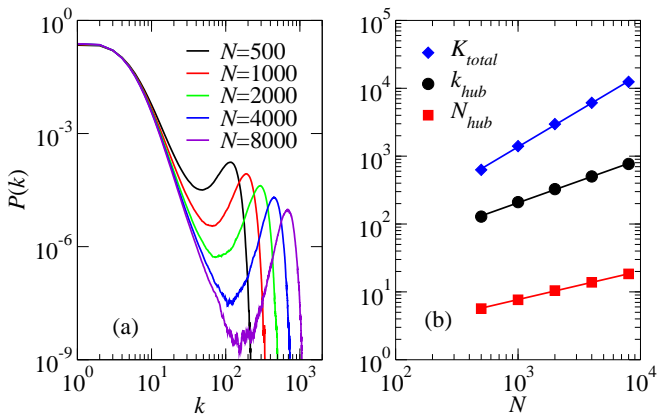}
\caption{Degree distribution in (a) and  $k_{hub}$, $N_{hub}$, and $K_{total}$
in (b) in the stationary state in systems with $\rho = 1.0$ and $r=0.01$.
The straight lines in (b) are guides to eyes.}\label{fig5}
\end{figure}

The model displays an interesting scaling behavior in the heterogeneous 
phase. Figure~\ref{fig5}(a) shows the degree distribution at several
values of $N$ to a given value of $\rho=1.0$ and $r=0.01$ belonging to the
heterogeneous phase. There is a peak corresponding to the hubs. 
As $N$ increases, the peak shifts to the right but does not sharpen nor
broaden.
This suggests that the number of hubs scales algebraically with $N$ and that
their degrees have the same order of magnitude scaling algebraically with
$N$. One can measure the number of hubs 
$N_{hub}$ from the spectral weight of the peak in $P(k)$. The numerical data
for $N_{hub}$ are plotted in Fig.~\ref{fig5}(b), which shows that it
follows a power law
\begin{equation}
N_{hub} \sim N^q
\end{equation}
with $q\simeq 0.43$.
Figure~\ref{fig5}(b) also shows that the maximum degree $k_{max}$ 
among all nodes
follows a power law
\begin{equation}
k_{max} \sim N^{\theta}
\end{equation}
with $\theta \simeq 0.64$. The sum of the exponents is close to unity, which
indicates that the total number of edges $K_{total}$ attached to the hubs
is proportional to $N$. Numerical data in Fig.~\ref{fig5}(b) shows that 
\begin{equation}
K_{total} \sim N^\mu
\end{equation}
with $\mu \simeq 1.07$ which is very close to 1. The exponents $q$,
$\theta$, and $\mu$ remain constant in heterogeneous phase
up to statistical errors.

The structural phase transition has a similarity to a condensation
transition in the zero range process~(ZRP)~\cite{Evans00}.
In the ZRP, a unit mass hops from one site to another on a given graph with
a hopping rate depending on the total masses on a departing site. 
The masses may undergo a condensation transition between a fluid phase and a
condensed phase. In the fluid phase, masses are distributed uniformly. 
When there is a strong on-site attraction among masses, a quenched disorder, 
or a structural heterogeneity in an underlying graph, a finite fraction of 
the total masses can condense on a single site to form a macroscopic 
condensate~\cite{Evans00,Hanney05,Noh05,Noh07}.

The similarity between the condensation in the ZRP and the emergence of 
hubs in network dynamics has already been noticed in Ref.~\cite{Angel05_07}.  
Regarding nodes and edges in the network dynamics as sites and masses in the
ZRP, respectively, a hub can be seen as a condensate of edges.
Apart from the similarity, the structural phase transition in our model has
a distinct feature. In the context of the condensation, there exist
multiple number of condensates~($N_{hub} \sim N^{q>0}$) and the condensates are
mesoscopic, that is to say, their size scales sublinearly in in
$N$~($k_{hub} \sim N^{\theta<1}$). This is contrasted to the ZRP with a 
local dynamics which has a single macroscopic condensate in the 
condensed phase~\cite{Evans00,Hanney05}.

Recent studies report that multiple
mesoscopic condensates appear when the dynamics is nonlocal in the sense that
the mass hopping rate depends not only on the local occupation
number but also on the global parameter  such as the system 
size~\cite{Angel05_07,Schwarzkopf08}. 
In our model, the edge rewiring dynamics
is purely local but coupled with the weight dynamics. We leave it as a
future work to understand the origin for the emergence of multiple
mesoscopic condensates in the context of the ZRP.

In summary, we propose a dynamic model for networks with the edge rewiring
dynamics which is coupled to the particle diffusion dynamics. The model
displays a structural phase transition between the homogeneous phase and the
inhomogeneous phase. The former is characterized by an exponential degree
distribution and the latter is characterized by the multiple 
mesoscopic hubs. Those hubs are the consequence of the positive feedback
between the edge and particle dynamics. Our work uncovers a mechanism how 
structural heterogeneity of complex networks can emerge.

This work was supported by Korea Research Council of Fundamental Science \&
Technology.

\end{document}